# Functional MRI applications for psychiatric disease subtyping: a review

Miranda L[1], Paul R[2], Pütz B[1], Müller-Myhsok B[1]

1: Statistical Genetics group of the Max Planck Institute of Psychiatry, Munich; 2: Department of Psychiatry and Psychotherapy, Section for Neurodiagnostic Applications, Ludwig-Maximilian University, Munich

**Acknowledgement of funding**

This project has received funding from the European Union's Horizon 2020 research and innovation programme under the Marie Sklodowska-Curie grant agreement No 813533.

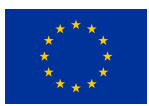

**Abstract**

*Background*

Psychiatric disorders have historically been classified using symptom information alone. With the advent of new technologies that allowed researchers to investigate brain mechanisms in a more direct manner, interest in not only the mechanistic rationale behind defined pathologies but also aetiology redefinition has greatly increased. This is particularly appealing for the field of personalised medicine, which searches for data-driven approaches to improve diagnosis, prognosis and treatment selection on an individual basis.

*Objective*

In the present article, we intend to systematically analyse the usage of functional MRI on both the elucidation of psychiatric disease biotypes and the interpretation/validation of subtypes obtained via unsupervised learning techniques applied to symptom or biomarker data.

*Methods*

Using PubMed, we searched the existing literature for functional MRI applications to the obtention or interpretation/validation of psychiatric disease subtypes in humans. The PRISMA guidelines were applied to filter the retrieved studies, and the active learning framework ASReviews was applied for article prioritization.

*Results*

From the 20 studies that met the inclusion criteria, 5 used functional MRI data to interpret symptom-derived disease clusters, 4 used it for the interpretation of clusters derived from biomarker data other than fMRI itself, and 11 applied clustering techniques to fMRI directly. Major depression disorder and schizophrenia were the two most studied pathologies (35% and 30% of the retrieved studies, respectively), followed by ADHD (15%), psychosis (10%), autism disorder (5%), and the consequences of early violence (5%). No trans-diagnostic studies were retrieved.

*Conclusions*

While interest in personalised medicine and data-driven disease subtyping is on the rise and psychiatry is not the exception, unsupervised analyses of functional MRI data are inconsistent to date, and much remains to be done in terms of gathering and centralising data, standardising pipelines and model validation, and method refinement. The usage of fMRI in the field of trans-diagnostic psychiatry, of great importance for the aforementioned goals, remains vastly unexplored.

**Introduction**

*Psychiatric disease prevalence*

Psychiatric disorders have a long history of being classified based solely on their associated symptoms, with the first attempts of systematic analysis dating back as far as 1840. Since the introduction of the Diagnostic Manual of Mental Disorders (DSM) back in 1952 [1], and most importantly since the inclusion of operationalized criteria in 1978 in the DSM-III [2], statistics on discrete pathological entities and their combination began to accumulate, yielding the potential of understanding psychiatric epidemiology in a consistent way. The last version of the DSM manual (DSM-5), published in 2013, contains 297 discrete disorders categorised into 11 broad classes, grouped together by evidence of co-occurring symptoms. Current prevalence estimates indicate that, on average, more than one in six (17.6%) have experienced at least one common psychiatric disorder within the last year, and almost three in ten (29.2%) during the course of their lifetime [3]. In an attempt to assess both the severity of the disorders and the response after individual treatment, several standardised symptom scores have been developed, including the Hamilton Depression Rating Scale (HAM-D) for Major Depression Disorder, and the Positive and Negative Syndrome Scale (PANSS) for Schizophrenia, among others.

*Heterogeneity and alternatives to symptom-based diagnosis*

Symptom and clinical information can be relatively easy to acquire, and their analysis can be useful to understand the symptom prevalence in the population and assess the effectiveness of treatment at a broad scale [4]. They do not, however, necessarily reflect anything about the underlying mechanisms causing those symptoms. Furthermore, given the complexity of the genetic and environmental factors at play, the same set of symptoms can arise from absolutely different causes, while the same biological causes may lead to different symptoms or phenotypes [5,6]. This is particularly important when analyzing the response to treatment, where the outcome is very difficult to predict based on the symptoms alone, and response to medication is vastly heterogeneous [7], being treatment-resistant variants of disease not uncommon. To name a few examples, current estimates indicate that about 30% and 34% of medicated patients diagnosed with depression and schizophrenia, respectively, do not respond to treatment even after trying two or more drugs [8,9]. This can be interpreted as an indication of the underlying mechanistic heterogeneity of these symptom-defined disorders. In light of this concern and with the advantage of new technologies and an increasing amount of related data, several initiatives have embarked on the quest of finding data-driven mechanistic disease definitions that may aid the issue. One of the most important to date has been the Research Domain Criteria (RDoC), which was introduced by the NIH in 2009 as a framework to guide research projects in the understanding of mental disorders from a combination of different perspectives, including not only self-reported symptoms but also genomics, circuits, and behaviour, among others [10]. The ideas behind these mechanistic-based classifications can not only expand our knowledge of mental disorders themselves and allow for biomarker-based classification but also aid in identifying the best treatments for individual patients whose overlapping symptoms have distinct aetiological causes, an idea that is very much in line with those of personalized medicine.

*Functional MRI for disease subtyping*

The idea of using multivariate pattern analysis to unravel the aforementioned heterogeneity, and unveil subgroups of patients within already defined diseases is not new [10–12]. However, the advent of massive biological related datasets (the so-called *high-throughput biology*) in areas such as genomics, transcriptomics, and proteomics and the newly available techniques to study the brain in a non-invasive way, opened a whole new field of possibilities to study not only the underlying mechanisms of symptom-related clusters but to search for biologically defined subtypes of disease (or *biotypes*) as well. Although initial hopes were put on mainly genetics, over the years an increasing number of GWA studies has revealed that brain disorders tend to be associated with a high number of genetic variants with very small effect sizes. Furthermore, individual genetic alterations often overlap among symptom-defined diseases [13]. While some progress in genetic biomarkers has been made using disease-specific polygenic risk scores (PRS), the usage of genetics alone for determining brain disease subtypes has been mostly elusive. One of the most promising fields to pursue this aim, however, has been that of neuroimaging, with Magnetic Resonance Imaging (MRI) as arguably its most proficient method to date. This technique has been increasingly used to study not only the structure of the brain (structural *MRI*) but also to measure changes in the blood oxygen levels surrounding particular regions, as a proxy of neuronal activation (*BOLD fMRI*) [14]. One of the most prevalent uses of this technology has been *task fMRI*, in which an experimental design matrix is typically convolved with a mathematical function modelling the haemodynamic response (called haemodynamic response function, or HRF) is set to explain the observed signal using a General Linear Model (GLM). While this approach has a substantial amount of literature behind and it is highly flexible due to relying on a Linear Model assumption [15], it has some notorious drawbacks. First, the most common analyses rely on what is called a *mass univariate test*, which statistically assesses differences in activation on each voxel separately, assuming independence even among contiguous regions in space. Second, it depends on a task experimental design, which even though it can be a powerful tool for answering specific questions is relatively hard to perform, difficult to generalise, and prone to habituation [16]. An alternative that gained momentum over the last two decades has been *resting-state* fMRI, in which no particular task is performed by the subjects. Since it was first employed in 1995 [17], this approach allowed researchers to study the relationship between brain regions over time, which has been proven to be a useful tool to study both *functional connectivity* (based on voxel correlation, yielding *undirected* connectivity networks) and *effective connectivity* (based on causal modelling, yielding *directed* connectivity networks). Regardless of the analysis tool, most studies largely converged in reporting multiple robust resting-state networks across the brain, such as the primary sensorimotor network, the primary visual network, frontoparietal attention networks and the well-studied default mode network [18]. Furthermore, the idea of the brain having stable connectivity between its different regions that can be altered in illness has been a powerful hypothesis for disease subtyping. Given its potential generalisability and the robustness of the obtained results [18,19], resting-state connectivity is currently the most used fMRI approach for both searching for and validating distinct mechanisms underlying brain disease, in an attempt of explaining the vast aforementioned heterogeneity.

*Unsupervised Machine Learning on psychiatric disease subtyping*

Automated pattern recognition can be used to unveil subtypes in psychiatric disease in an unsupervised way (that is, without the presence of hardcoded labels). Given the complexity of the data at play, this set of approaches has been proven extremely useful in a variety of settings and data domains not only for *clustering* but also for *dimensionality reduction*. While the former deals with the process of finding subtypes in itself, the latter encapsulates a set of methods to project the data into lower-dimensional manifolds while retaining most of its original information. In the case of functional MRI, unsupervised machine learning has been extensively important given the inherent absence of structure in the data. Its main uses include but are not restricted to **parcellation of the brain into discrete functional subunits** (unravelling of brain connectivity networks), the study of **brain connectivity dynamics** (how those networks evolve over time), and **grouping subjects according to their connectivity features** (used for disease subtyping in itself). The first two mentioned uses fall into the dimensionality reduction category; the third is inherent to clustering.

This review will analyse the reported use to date of fMRI for unveiling subtypes in several psychiatric disorders, as well as as a tool for validating subtypes reported in symptom scores and structural MRI. The strengths and weaknesses of each approach will be discussed. For a detailed review of the existing unsupervised learning methods for disease subtyping, see Marquand at al (2016) [20]. For methods on resting-state fMRI in particular, refer to Khosla et al (2019) [18].

## Materials and Methods

For consistency with previous work, this study followed the Preferred Reporting Items for Systematic reviews and Meta-Analysis (PRISMA) statements [21].

*Search methods for article retrieval*

A systematic search of original articles was carried out on the PubMed database, including all non-review articles from the date of database creation up to 25 May 2020. The string **"(unsupervised learning OR clustering OR dimensionality reduction OR subtyping) AND functional MRI"** was introduced on the search engine, with the intention of retrieving all available papers in which functional MRI was used either for brain disease subtyping or for validation of brain disease subtypes obtained via other methods, which should include at least one of symptom information and structural MRI data.

*Article filtering*

All retrieved studies were downloaded and analysed using PubMed metadata to filter review articles ("D016428:Journal Article", but  "D016454:Review" absent in the *'publication_types'* metadata field). The remaining studies were analysed using the ASReviews (Automatic Systematic Reviews) python package, an active learning based recommender system that trains a classifier on the abstracts of the provided papers, in order to present the user with the most relevant papers to review. While all abstracts included in this step were carefully studied, this tool has proven to be useful for prioritisation. Studies whose abstracts met the exclusion criteria (see below) were discarded. The rest was selected for full-text review.

*Inclusion/exclusion criteria*

All original non-review studies in which functional MRI was used either for brain disease subtyping directly or for validation of brain disease subtypes obtained via other methods, which should include at least one of symptom information and structural MRI data. Disease subtyping had to be carried out in a fully unsupervised way (no labels based on prior information). Clear definitions of the methods and their validation had to be included. As, given the heterogeneity of results, we think that cluster validation is currently one of the most important discussion topics in the field, articles trying to replicate or validate the results of included studies were also included.

*Data extraction for systematic analysis*

For each article that was included in the final review, a set of systematically collected pieces of information was extracted and added as an entry to a table (see tables 1, 2, 3). This information includes: (a) *Publication year*, (b) Title, (c) *Implicated brain disease*, (d) *sample size*, (e) presence of preprocessing/dimensionality reduction, (f) *clustering technique employed*, (g) *cluster-number selection procedure*, (h) *Healthy Controls included in the subtyping procedure*, (i) *presence-of-clusters statistical testing (against continuum)*, and (j) *data employed*[1].

---

[1] Given that fMRI data can be used for either validation of symptom clusters, validation of biomarker clusters or subtyping itself.

## Results

A total of 144 related articles were retrieved from PubMed in the first place, of which 120 were retained after filtering for duplicated studies and reviews. 2 studies identified through manual search were also included, yielding a total of 122 articles selected for abstract inspection. During the aforementioned ASReview assisted procedure, a total of 35 articles were selected for a full inspection. The final number of studies that met the inclusion criteria was 20. A scheme of the full pipeline is shown in Figure 1.

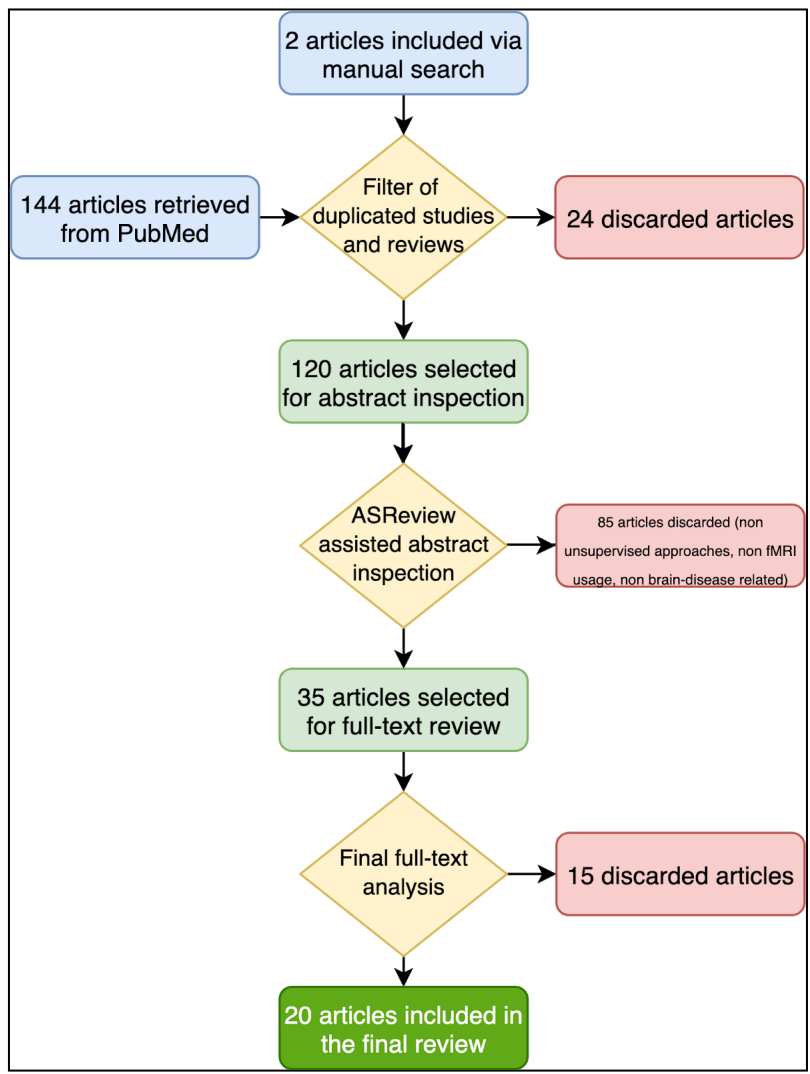

**Fig.1.** PRISMA flowchart that schematizes the employed pipeline. A total of 144 articles were retrieved from PubMed using the string "*(unsupervised learning OR clustering OR dimensionality reduction OR subtyping) AND functional MRI*", and 2 articles were included after manual search, yielding a total of 146 input studies. After several systematic filters, which included exclusion of reviews, duplicated articles and relevance to the defined inclusion criteria, a total of 20 articles was included in the review.

Light blue nodes indicate article input. Yellow nodes indicate filtering procedures. Light red nodes indicate discarded articles. Light green nodes indicated accepted articles during intermediate steps. The final green node indicates the accepted articles after the implementation of the whole pipeline.

*Characteristics of the included studies*

As previously mentioned, 20 full-articles were included in this review after the implementation of the specified pipeline. These studies were classified into one of three categories based on the nature of the analysed subtypes and the usage of functional MRI (Figure 2). The classes are: (a) *fMRI used for validation of subtypes obtained via unsupervised learning of symptom-related data*, (b) *fMRI used for validation of subtypes obtained via unsupervised learning of biomarkers other than fMRI (including structural MRI)*, and (c) *fMRI used for brain disease subtyping directly*. Over the next three sections we will analyse these three cases separately, summarising the results that the respective studies report, and discussing the assumptions they make and the advantages and disadvantages that they imply.

Regarding the pathological entities under study, the majority of the articles analysed patients diagnosed with **Major Depression Disorder** and **Schizophrenia** (38.1% and 28.6%, respectively. **Psychosis**, **Attention Deficit Hyperactivity Disorder**, **Autism Disorder** and consequences of **early violence** were also included.

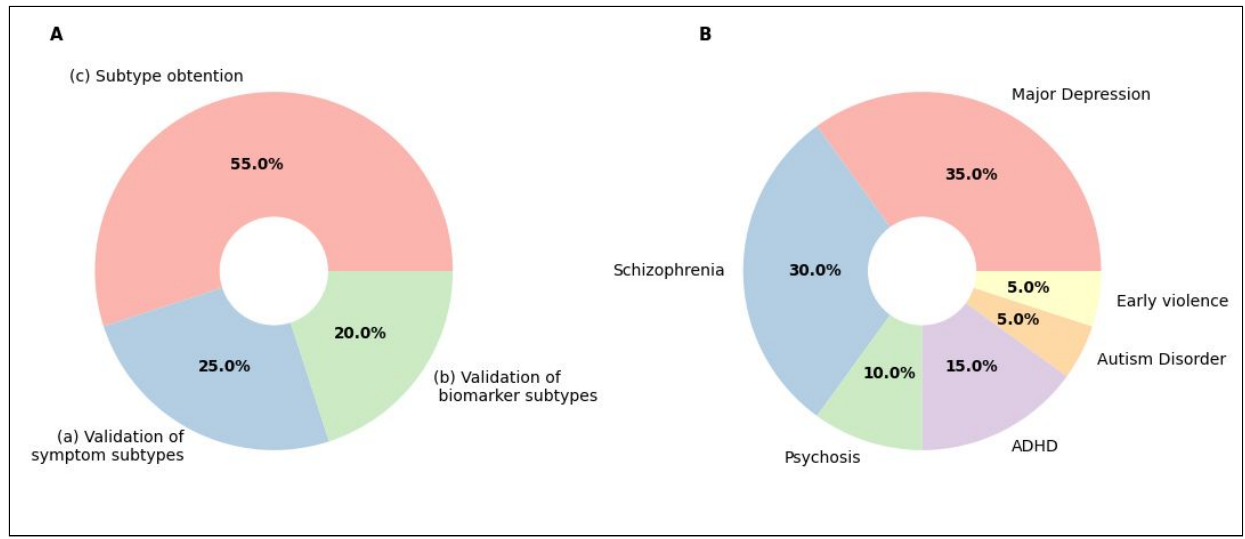

**Fig.2. (A)** Doughnut plot representing the number of selected studies for each of the three defined categories: (a) *fMRI used for validation of subtypes obtained via unsupervised learning of symptom-related data*, (b) *fMRI used for validation of subtypes obtained via unsupervised learning of biomarkers other than fMRI (including structural MRI)*, and (c) *fMRI used for brain disease subtyping directly.* **(B)** Doughnut plot representing the most prevalent brain disorders that the included studies analysed.

*fMRI used for validation of subtypes obtained via unsupervised learning of symptom-related data*

The first category to analyse corresponds to articles in which clusters were defined from symptom-based scales and functional MRI was employed as a tool for cluster interpretation and/or validation.

The unsupervised classification of psychiatric symptoms is not new: to our knowledge the first papers were published back in the 1970s [10–12]. The novelty of the studies presented here relies on the efforts for interpreting and validating symptom clusters in terms of functional mechanisms. By comparing functional MRI data coming from patients on different clusters, researchers can potentially explain which mechanisms may be at play when yielding distinct sets of symptoms. After the systematic screening, five original studies covering both Major Depression Disorder and Schizophrenia fell into this section. In the following paragraphs, they will be discussed, ordered by disease.

Taubner et al [22], analysed a small cohort of 20 subjects using Q factor analysis on SWAP-200 [23] symptom data. Using an elbow method on the variance explained by PCA, they fit a two-factor model that yielded two distinct patient types, defined by depressive personality and emotional-hostile externalising personality. A whole-brain correlation analysis was performed between the loadings of each factor and the GLM betas fitted to fMRI data, obtained during a stress-related task. While no correlation was significant for factor 1, factor 2 yielded significant results in four regions: the right orbitofrontal cortex, the left ventral striatum, the left temporal pole and the right middle frontal gyrus. While the sample size is small and the proceedings arguably simple, the study calls itself a hypothesis-generating experiment that might be followed up in the future.

A different approach, with bigger sample size, was employed by Maglanoc et al in 2018 [24]. Using a subspace clustering variant of Gaussian Mixture Models, the authors found five clusters of both major depression patients and healthy controls in data coming from Beck's depression and Beck's anxiety inventories. They explored the relationship between symptom occurrence in each cluster and both static and dynamic functional connectivity matrices obtained from resting-state fMRI using ANCOVA. While there were no significant results from the dynamic connectivity analysis, there were significant correlations between the most severe depression cluster and the frontotemporal network.

The rest of the studies in this section focused on Schizophrenia. The first, published by Geisler et al. in 2015 [24,25], aims to obtain subtypes not from symptoms directly but from a behavioural and cognitive score that they claim to be more longitudinally stable than standard symptom collections. Using a sample size of 129 patients and 165 healthy controls (not used for clustering), they derived four clusters using K-means on an 8 dimensional PCA obtained from the behavioural and cognitive data. Both the number of components for PCA and the number of clusters for K-means were obtained using an elbow method on the variance explained and inertia curves, respectively. The clusters were characterised by a specific pattern of structural brain changes in areas such as Wernicke's area, lingual gyrus and occipital face area, and hippocampus as well as differences in working memory-elicited neural activity in several frontoparietal brain regions. The latter was assessed using functional MRI obtained during a blocked SIRP task.

In 2018, Dickinson and colleagues published an article [26] in which they attempted to take a different approach, by clustering data coming from the Positive And Negative Syndrome Score (PANSS), a widely-used standardised schizophrenia-specific symptom scale. Using a sample size of 549 individuals comprising only diagnosed patients and the 2-step SPSS clustering algorithm, they obtained three clusters that were characterised as deficit (with enduring negative symptoms and diminished emotionality), distress (with high emotionality, anxiety, depression and stress sensitivity) and low-symptomatic. A subsample of 182 patients was exposed to functional MRI scans during a working memory task. When comparing the activations between the already defined clusters, statistically significant differences emerged in primary regions of the frontoparietal working memory network including right dlPFC, left anterior cingulate and left parietal cortex. The low-symptom subgroup showed significantly greater activation in the right dlPFC during working memory performance than the two more symptomatic groups, a "healthier" pattern of prefrontal engagement during the performance.

**Table 1:** Retrieved articles in which fMRI was used to interpret symptom-based clusters

| Publication year | Title | Pathology | sample size (symptoms) | sample size (fMRI) | data | Preprocessing / Dimensionality reduction | Clustering | Model selection | Healthy Controls included | Stability testing | Continuum testing | Subtypes found |
|---|---|---|---|---|---|---|---|---|---|---|---|---|
| 2013 | Neural activity in relation to empirically derived personality syndromes in depression using a psychodynamic fMRI paradigm | Major Depression | 20 | 20 | SWAP-200, task fMRI | Whole brain correlation between task and GLM betas | Q-Factor analysis | PCA | No | Not specified | No | 2 |
| 2015 | Brain structure and function correlates of cognitive subtypes in schizophrenia | Schizophrenia | 294 | 294 | longitudinally stable cognitive scales, sMRI, fMRI | PCA | K-means | Inertia elbow method | Yes | Not specified | Yes (in model selection) | 4 |
| 2018 | Attacking Heterogeneity in Schizophrenia by Deriving Clinical Subgroups From Widely Available Symptom Data | Schizophrenia | 549 | 182 | PANSS scale, task fMRI | No | SPSS 2-step clustering | BIC | No | Bootstrap | Yes (in model selection) | 3 |
| 2018 | Data-driven clustering reveals a link between symptoms and functional brain connectivity in depression | Major Depression | 1084 | 251 | BDI, rs fMRI | No | subspace GMM | BIC | Yes | Jaccard index stability | Yes (in model selection) | 5 |
| 2020 | Neurobiological divergence of the Positive and Negative Schizophrenia subtypes identified on a new factor structure of psychopathology using non-negative matrix factorization: an international Machine Learning study | Schizophrenia | 1545 | 84 | PANSS, rs fMRI | Non-negative Matrix Factorization | fuzzy-c means / GMM | fuzzy silhouette index / BIC | No | leave-one-site out CV, bootstrap | Yes (in model selection) | 4 |

A similar approach was followed by Chen et al. in 2020 [26,27]. Using a bigger sample of 1545 patients diagnosed with schizophrenia, they used Non-Negative Matrix Factorization to reduce the dimensionality of patients' PANSS score data. The optimal value of four factors was obtained using cross-validation after 10.000 split-half runs on the Rand index, VOI (variation of information index) and the concordance index between dictionaries. These factor loadings were used as input for clustering using the fuzzy-c-means algorithm, assisted by Gaussian Mixture Modelling to delimit the threshold between clusters.

The optimal cluster number was determined using the fuzzy silhouette index among others, and cluster stability was addressed via leave-one-site out replication, subsampling and bootstrap. Regarding MRI, functional undirected connectivity was obtained from resting-state data, and the extracted networks were used to predict cluster membership in a supervised manner using an RBF Support Vector Machine. A feature importance analysis of this classifier yielded profiles of the ventromedial frontal cortex, temporoparietal junction, and precuneus as the most important networks for cluster assignment.

*fMRI used for validation of subtypes obtained via unsupervised learning of biomarker data*

In this second section, we will discuss three studies (published across four papers) in which the obtention of biotypes was attempted applying unsupervised learning techniques to sets of biomarkers other than functional MRI itself. Among the included biomarkers, one of particular interest is structural MRI, for which the necessary assumptions underlying structure and function will be discussed in the following paragraphs.

The first study is composed of two articles, published by Clementz et al and Meda et al in 2015 and 2016 [28],[29] on the identification of psychosis biotypes. The first article deals with the objection of the biotypes themselves, the second analyses their functional correlates using resting-state functional connectivity.

The term psychosis refers to the general concept applied to several pathologies that lead to a deteriorated perception of reality. The authors of this study claim that there may be different aetiologies underlying psychotic symptoms that do not necessarily overlap with the symptom-defined labels available (schizophrenia, schizoaffective disorder, and bipolar disorder with psychosis). They gathered 1872 samples from patients diagnosed with any of these diseases (n=711), their first-degree relatives (n=883) and comparable healthy subjects (n=278). The data consisted of biomarker panels comprising neuropsychological markers, stop-signal data, saccadic control data and auditory stimulation paradigms. Patient data were used for clustering, while relatives and controls served for result interpretation. The data were first processed using the pre-clustering step of the SPSS 2-step clustering algorithm to obtain a nine-dimensional feature space that they fed into a K-means algorithm for finding the biotypes. The number of clusters was selected using the gap statistic, yielding a three-component solution that did not overlap with the DSM-5 defined labels mentioned above. They hypothesized that these three components may correspond to three different aetiologies, that over the patients' lifespan lead to similar sets of symptoms, and observed that clusters differed in outcome severity, cognition measures and sensorimotor reactivity, among others.

In the follow-up study, individuals in an independent sample were assigned to the already defined clusters, and comparisons between their ICA obtained functional resting-state connectivity profiles were made by means of linear mixed models for within-proband differences across categories and correlations with the aforementioned biological profile scores. They claim that their biotypes performed marginally better in terms of separating out psychosis sub-groups from their functional connectivity data compared to conventional DSM-5 diagnosis. When comparing patients to relatives and healthy controls, they found significantly reduced connectivity on specific biotypes in nine networks, including the cuneus-occipital network, the left and right FPN networks, the cerebellar occipital network, the anterior, IP and SP default mode networks, the temporoparietal network, and the frontoparietal control network. All these deficits are claimed to track more closely with cognitive control factors, suggesting potential implications for both disease profiling and therapeutic intervention.

The remaining two studies, published by Chen et al in 2019 [30] and Kaczkurkin et al in 2020 [31], used structural MRI to find disease subtypes and projected their findings into resting-state functional connectivity data afterwards.

The first of the two attempts to find Autism Disorder subtypes in a sample of 759 individuals comprising both patients and healthy controls. After using Non-negative Matrix Factorization to reduce the dimensionality of a Voxel-Based Morphometry analysis of the structural data, clustering was performed using the K-means algorithm. Using the silhouette index for model selection, the analysis yielded a three-component solution, which they claim confirms that ASD is not a neuroanatomically homogeneous disease. When comparing the obtained clusters with both symptomatic scores, they found that clusters showed differences in disease severity. When comparing the resting-state functional connectivity networks obtained with the *data processing assistant for resting-state fMRI* tool (DPARSF) [32] in each cluster to healthy controls, they found statistically significant differences in two of the clusters. ASD patients had diminished connectivity in the default mode network, the frontoparietal network, the cingulo-opercular network and the occipital network.

The second work, focused on finding structural subtypes in subjects with internalising disorders, takes a particular approach to disease subtyping. Instead of clustering in a fully unsupervised way, the authors use a semi-supervised approach called HYDRA [33] which uses the binary disease-control labels to find different disease subtypes regarding their difference to controls. To achieve this they employ a multiple linear SVM classifier under the hood that both maximizes the margin between cases and controls for each cluster and the margin between clusters. Using this algorithm in volumetric and cortical thickness data coming from 1141 individuals, they found a two disease-cluster solution when maximising the adjusted Rand index (ARI) during cross-validation. When analysing both structural particularities of the subject belonging to each cluster, as well as symptomatic and cognitive measures, the authors found that one of the subtypes showed smaller volume, thinner cortex, reduced white matter integrity, greater psychopathology and poorer cognitive performance than its counterpart. The functional connectivity of 40 subjects assigned to these two defined categories was obtained in the frequency space, by computing the voxelwise amplitude of the low-frequency (0.01-0.08 Hz) band of the power spectrum (ALFF). This approach has the advantage of allowing the direct comparison of structural and functional measures using the same atlas [34], the functional measures being a reflection of the average connectivity of a particular region of interest, in this case, delimited by differential structural measures. Using this approach, the authors found significant differences in average connectivity between clusters in frontal regions, the right amygdala and the right hippocampus, which correlated with poorer function across multiple domains. They claim that the identification of biologically grounded internalising subtypes may assist in targeting early interventions and assessing longitudinal prognosis.

**Table 2:** Retrieved articles in which fMRI was used to interpret biomarker-based clusters[2]

| Publication year | Title | Pathology | sample size (biomarkers) | sample size (fMRI) | data | Preprocessing / Dimensionality reduction | Clustering | Model selection | Healthy Controls included in clustering | Stability testing | Continuum testing | Subtypes found |
|---|---|---|---|---|---|---|---|---|---|---|---|---|
| 2015 | Identification of Distinct Psychosis Biotypes Using Brain-Based Biomarkers | Psychosis | 1872 | - | biomarkers | SPSS pre clustering | k-means | gap statistic | No | No | Yes (in model selection) | 3 |
| 2016 | Examining Functional Resting State Connectivity in Psychosis and its subgroups | Psychosis | 1125 | 1125 | symptoms, rs fMRI | Functional connectivity (ICA) | - | - | - | - | - | - |
| 2019 | Parsing brain heterogeneity in males with autism spectrum disorder reveals distinct clinical subtypes | Autism Disorder | 759 | 403 | sMRI, symptoms, fMRI | Non-negative Matrix Factorization | K-means | silhouette index | No | Bootstrap | No | 3 |
| 2020 | Neurostructural Heterogeneity in Youths With Internalizing Symptoms | Internalizing disorders | 1141 | 40 | sMRI, symptoms, fMRI | knowledge-based feature selection | HYDRA | Stability based (Adjusted Rand Index) | Yes | Adjusted Rand Index, cross validation | No | 2 |

---

[2] The first two articles in the table are complementary. Whilst the first provides the biomarker clustering, the second contributes with a functional MRI interpretation of the same analysis, hence they are both included.

*fMRI used for brain disease subtyping directly*

The last results section will deal with studies in which biotype obtention was attempted from functional MRI data itself. Eleven articles (ten original studies and a very relevant replication) comprising four disorders were included, ten of which relied on resting-state functional or effective connectivity. The implications of each methodology will be discussed after an overview of each individual work.

Starting with Schizophrenia and its related disorders, in 2014 Yuhui Du et al published an article in which the distinction between Schizophrenia, Schizoaffective disorder and psychotic bipolar disorder, all of which share a common set of symptoms, was attempted to be redrawn using functional connectivity data clustering [35]. They employ Group Information Guided ICA (GIG-ICA) [36] to obtain subject-specific functional connectivity networks from a sample of 93 individuals containing patients diagnosed with any of the mentioned pathologies as well as healthy controls. For each subject, a vector was constructed with the connectivity features that remained after an SVM recursive feature elimination procedure [37], and an interindividual distance matrix was constructed using the inverse of Pearson's correlation coefficient. All three of K-means, n-cut and Hierarchical Agglomerative Clustering were used with K=5, and the accuracy in retrieving the original groups was assessed. During feature selection, regions such as the default mode network, frontoparietal networks, salience networks, auditory-related network, parietal network, vision and visuospatial networks, cerebellum and sensory-motor networks were prioritised, indicating that the connectivity among these regions may be of interest for distinguishing between the pathologies at hand. While this study did not delve into the discovery of new subtypes of a predefined disease, it provided connectivity evidence for the distinction and validation of distinct disease entities whose symptoms overlap, adding fresh evidence to an ongoing discussion in the field.

Another article that dealt with dissecting the mechanistic underpinnings of Schizophrenia and its potential subtypes was published by Brodersen et al in 2014 [38]. In this proof-of-concept study, the authors employed a dynamic causal modelling algorithm to retrieve a directed connectivity model from a sample of 83 subjects including diagnosed patients and healthy controls. By employing a variational Bayesian variant of a Gaussian Mixture Model-based clustering algorithm in the DCM parameters, the authors were able to retrieve a three cluster solution from the patients alone, whose components significantly differed in their clinical manifestation, as assessed by the aforementioned PANSS scale. When including the healthy controls, however, the best-found solution (the one with the lowest BIC score) had only two components that overlapped ~72% with the case/control labels. The authors use these results as an argument to defend the exclusion of healthy controls in the unsupervised learning procedure, as the likelihood of the already-known binary factor is high.

The last covered article on (early onset) Schizophrenia subtyping was published by Yang et al, also in 2014 [39]. Using a small sample of 52 individuals including both diagnosed patients and healthy controls, the authors used a pipeline called gRACIAR (*generalized ranking and averaging independent component analysis by reproducibility*) [40] to obtain both subject-specific functional connectivity networks and a meta graph concerning intersubject similarity. Using Newmann's community detection algorithm with cross-validated binarizing thresholds, they obtained a two-component solution whose clusters were differentiable by means of the PANSS scale. Furthermore, a feature importance analysis revealed the crucial importance of the precuneus angular gyri, the superior temporal gyri and the inferior frontal gyri for early-onset schizophrenia biomarking.

Drysdale et al in 2017 [41] used a Canonical Correlation Analysis (CCA) to reduce the dimensionality of resting-state functional connectivity data coming from a sample of 312 patients diagnosed with MDD, in a way that maximized its correlation with symptom data coming from HAM-D scores. While this approach can be useful for integrating the two types of data, conclusions derived from it must be taken with caution: a transformation of the biological data that correlates with the final phenotypic outcome could shed light into biological markers to identify those phenotypes in a less subjective way than a questionnaire, but their correspondence to biotypes or distinct aetiologies, as previously mentioned, can be questioned. The authors then used the first two canonical variates obtained from CCA (which they interpreted as anhedonia and anxiety-related) to cluster individuals and reached a four-component solution using a Hierarchical Agglomerative Clustering approach. When projecting the obtained clusters into the original resting-state connectivity data, differential activations were observed in the limbic and frontostriatal networks. In addition, they reported that cluster membership accurately predicted treatment response, as assessed via HAM-D evaluation during and after Transcranial Magnetic Stimulation treatment.

While ambitious and inspiring, this article received strong criticism in a replication attempt made by Dinga et al in 2019 [42]. While following nearly the same pipeline on a smaller independent cohort of 187 individuals, the authors highlighted several statistical weaknesses in the original study. First, they claim that there is a bias in the statistical testing of the CCA results in the article published by Drysdale et al. While the original article reports that both canonical variates' correlation with symptoms is statistically higher than random, permutation testing in the replication study showed that this seems not to be the case when taking into account the pre-selection of voxels that were most correlated with the symptoms themselves. Dinga et al claim that the original procedure is likely to be selecting noise in the direction of the hypothesis, and not correcting the consequent statistical tests afterwards.

**Table 3:** Retrieved articles in which fMRI was used to cluster subjects into biotypes

| Publication year | Title | Pathology | sample size (fMRI) | data | Preprocessing / Dimensionality reduction | Clustering | Model selection | Healthy Controls included | Stability testing | Continuum testing | Subtypes found |
|---|---|---|---|---|---|---|---|---|---|---|---|
| 2014 | Exploring difference and overlap between schizophrenia, schizoaffective and bipolar disorders using resting-state brain functional networks | Schizophrenia, Schizoaffective disorder, bipolar disorder | 93 | rs fMRI | Functional connectivity (GIG-ICA) | k-means, n-cut, HAC | No | Yes | No | No | 5 |
| 2014 | Dissecting psychiatric spectrum disorders by generative embedding | Schizophrenia | 83 | rs fMRI | Effective connectivity (DCM) | Variational Bayesian GMM (on connectivity parameters) | BIC | Yes/No | No | Yes (in model selection) | 2/3 |
| 2014 | Organizing Heterogeneous Samples Using Community Detection of GIMME-Derived Resting State Functional Networks | ADHD | 80 | rs fMRI | Effective connectivity (GIMME) | Newmann's Graph-based Community Detection | Reachability-based thresholding | Yes | Bootstrap | No | 5 |
| 2014 | Brain network informed subject community detection in early-onset schizophrenia | Schizophrenia | 52 | rs fMRI | Functional connectivity (gRACIAR) | Newmann's Graph-based Community Detection | Reachability-based thresholding | Yes | Bootstrap | No | 2 |
| 2015 | Characterizing heterogeneity in children with and without ADHD based on reward system connectivity | ADHD | 106 | rs fMRI | Neurosynth meta analysis, Functional connectivity | Newmann's Graph-based Community Detection | Reachability-based thresholding | Yes | Permutation VOI | No | 3 |
| 2017 | Resting-state connectivity biomarkers define neurophysiological subtypes of depression | Major Depression | 312 | rs fMRI, HAM-D | Canonical Correlation Analysis | Hierarchical Agglomerative Clustering | Maximum ratio of between-cluster to within-cluster variance | No | No | No | 4 |
| 2017 | Data-Driven Subgroups in Depression Derived from Directed Functional Connectivity Paths at Rest | Major Depression | 80 | rs fMRI | Effective connectivity (GIMME) | Walktrap Graph-based Community Detection | Handled by the clustering algorithm | No | Based on random perturbations of the similarity matrix | No | 2 |
| 2018 | Brain-behaviour patterns define a dimensional biotype in medication-naive adults with attention-deficit hyperactivity disorder | ADHD | 203 | rsfMRI, behavioural data | Canonical Correlation Analysis (NBS, behavioural data) | k-means/ spectral clustering | Silhouette, Jacquard, GAP / similarity threshold perturbation | Yes | No | Yes | 1 |
| 2018 | Identification of depression subtypes and relevant brain regions using a data-driven approach | Major Depression | 134 | rsfMRI, BDI, SNPs, Methylation, biomarkers | Functional connectivity | Multiview co-clustering | Best view in separating cases and controls (Cohen's D) | Yes | Multiple random initializations | No | 5 |
| 2019 | Evaluating the evidence for biotypes of depression: Methodological replication and extension of Drysdale et al. (2017) | Major Depression | 187 | rs fMRI, IDS | Canonical Correlation Analysis | Hierarchical Agglomerative Clustering | - | No | Jackknife analysis | Yes | 1 |
| 2020 | Biotypes of functional brain engagement during emotion processing differentiate heterogeneity in internalizing symptoms and interpersonal violence histories among adolescent girls | Early violence | 114 | task fMRI | Neurosynth meta analysis, GLM parameters | k-means | Inertia elbow method | Yes | Jackknife analysis | Yes (in model selection) | 3 |

In addition, they mention that even when doing a proper model selection procedure, the original study does not test the hypothesis that there is an inherent clustering structure in the data, against the possibility of a continuum. When testing this using previously described methods [43], they find no significative evidence supporting clustering. While some details of the proceedings were not the same, such as the symptom scale used or some differences in fMRI preprocessing, this article shows how important proper statistical testing is in these complex scenarios of multiple data integration and how crucial replication attempts are.

While all studies mentioned in this section so far dealt with data coming from resting-state *functional* (and thus undirected) connectivity, Price et al were the first to our knowledge, in 2017, to use *effective* connectivity to build directed resting-state networks using causal modelling for brain disease subtyping [44]. The algorithm employed for graph construction (called Group Iterative Multiple Model Estimation, or GIMME) has been extensively shown to reliably recover both the presence and direction of connectivity among brain regions per individual in simulations. Using a sample of 80 diagnosed patients with Major Depression, the authors built a meta-network based on the correlations between model parameters among individuals, where the linking thresholds were controlled using statistical testing based on network random perturbation. Applying a Walktrap community detection algorithm in this meta-graph yielded a two-component solution, whose stability was tested by perturbing the graph as well. The biggest group showed a typical connectivity pattern across DMN nodes, as previously reported on average depressed patients. The smaller group, however, showed atypical connectivity in this region, with increased dorsal anterior cingulate-driven connectivity paths. This smaller group had also significantly higher comorbidity with an anxiety disorder and highly recurrent depression, which lead to a poorer outcome of the disorder. While the employed sample size is small, this study illustrates how graph theory and causal modelling can be used together to shed light into the mechanisms behind major depression in particular and brain disorders in general.

Continuing with MDD, a very different unsupervised approach was taken by Tokuda et al in 2018 [45]. The authors of this article employed a custom made co-clustering algorithm capable of grouping both features and subjects, yielding multiple subspace solutions, or *views*, that can explain multiple grouping structures present in the data. While some might be irrelevant to the problem at hand (such as age or sex), by selecting the *view* that correlates the most with pre-accepted labels (e.g. that separates cases and controls the most accurately) it is possible to retrieve a relevant clustering solution and feature selection on the same step. Furthermore, this method allows the authors to integrate resting-state functional connectivity data with other data domains, such as BDI questionnaires, biomarker panels, and genetics and methylation data coming from a preselected set of related genes. They used this approach to cluster a sample of 134 subjects including both patients and controls and selected the *view* with the greatest Cohen's D coefficient when separating cases and controls. The algorithm yielded a five-component solution, of which two corresponded to controls and three to patients almost exclusively. The three MDD-related reported clusters were observed to differ significantly by functional connectivity between the Angular Gyrus and other brain areas in default mode networks, child abuse trauma scale scores (CATS) and selective serotonin reuptake inhibitor treatment outcomes. While the employed sample size is relatively small and the results demand replication, this article proposes an innovative and powerful approach with a high potential for integrating distinct data domains.

The first article to mention attention deficit hyperactivity disorder (ADHD) was published in 2014 by Gates et al [46]. Using the aforementioned GIMME algorithm to recover directed connectivity graphs from a sample of 80 individuals including both diagnosed patients and healthy controls, the authors applied Newmann's community detection to a meta-network based on intersubject connectivity correlation to retrieve a solution with five components whose stability was evaluated via bootstrapping. While the obtained subgroups are highly distinguishable by their differential connectivity in regions such as the dorsolateral prefrontal cortex, the frontal cortex, the intraparietal sulcus, and the inferior parietal lobule, the presence of both cases and controls in each of the clusters makes it difficult to analyse how clinically relevant the reported structure can be. The study, however, serves as a proof of concept for the highly standardised pipeline it proposes and opens the discussion for further exploration.

Using a functional connectivity pipeline on a sample of 106 children (aged 7-12 years) including both diagnosed patients and controls, Costa Dias et al also attempted to find data-driven subtypes of ADHD in their article published in 2015 [47]. By means of a meta-analytic mask obtained from NeuroSynth [48] which was centred on the reward system, the authors filtered the resting-state functional MRI time-series before feeding them to a correlation-based algorithm for connectivity extraction. A meta correlation matrix was obtained to assess interindividual similarities, and a three cluster solution was obtained using Newmann's community detection, which was stability tested by means of the Variation of Information (VOI index) under random network permutations. While all three subgroups contained both cases and controls, there were significant community-specific connectivity differences between patients and healthy subjects. Furthermore, impulsivity-related behavioural scores were significantly distinct between cases and controls only in one of the subgroups, indicating that the retrieved functional subgroups might be related to specific behavioural characteristics.

A very different conclusion was reached by the last included article on ADHD, published by Lin et al in 2018 [49]. The authors criticize previous efforts for both the inclusion of healthy controls in the clustering sample and the lack of testing for the presence of a subgroup structure before applying the unsupervised learning techniques. By contrasting what they called a dimensional biotype (a non-discrete severity continuum of the same pathological entity) against a categorical biotype, they claim that they lack evidence for the latter and that therefore clustering structures previously reported should be interpreted with caution. The methods they employ to reach these statements, however, are very different from the ones described before, making direct comparisons difficult. They obtain functional connectivity networks using an ICA based pipeline and compute the Network-Based statistic for every diagnosed patient (n=80) and matched control (n=123). Next, they compute the canonical correlations (CCA) between the NBS results and a set of symptoms assessed by factor scores of inattentive and hyperactivity-impulsivity symptoms and FIQ. When assessing the clustering structure both via K-means of the CCA results and spectral clustering of the NBS and the symptoms together, they do not find significant evidence backing more than one cluster (as assessed by the average silhouette index, the Jaccard similarity index and the GAP statistic, and the convergence of the number of clusters across different similarity thresholds, respectively). While their conclusions are valid when including the symptoms in the clustering process (either directly or indirectly via CCA), the analysis does not rule out, in our opinion, the possibility of several independent functional entities behind a unique set of symptoms that can be interpreted as a continuum, and more research is needed to solve this issue.

The last paper in this section, published by Sellnow et al in 2020 [50], delves into the functional consequences of violence in early childhood and internalising symptoms. Using a sample of 114 adolescent girls (aged 11-17), the authors used functional MRI obtained during an emotion processing task in a blocked design. After filtering the voxels of interest using a meta-analytic mask obtained from NeuroSynth (related to emotion processing), the GLM first order betas were clustered using the K-means algorithm. After selecting the best model in terms of the elbow method on the cluster validity index they reached a three-component solution, proven stable via leave-one-out cross-validation. The clusters were distinguishable by engagement of the medial prefrontal cortex, the anterior insula, the hippocampus, and the parietal and ventral visual cortex during emotion processing. When analysing the relationship between each cluster and measures of interpersonal violence (IPV) and internalised symptoms, the authors reported differential correlations per group. Furthermore, as IPV exposed a negative correlation with symptom reduction over Trauma-Focused Cognitive Behavioural Therapy (TFCBT), the authors explored the possibility of their methodology to predict treatment outcome based on functional information.

**Discussion**

This article reviewed the existent methodologies and approaches to employ functional MRI data in precision psychiatry, and more precisely in brain disease data-driven subtyping. A total of twenty articles were retrieved from the PubMed database after a systematic screening, in which fMRI was used either for validation/interpretation of subtypes obtained from symptom or biomarker data clustering or for biotype obtention itself.

While the first section of the results provided examples that illustrate how powerful unsupervised learning can be to detect subgroups in psychiatric symptom data (table 1), it's important to keep in mind several fundamental facts when using these tools. First, as discussed in the introduction, different sets of symptoms do not necessarily reflect distinct aetiologies. Second, symptoms are sensitive to treatment and environmental perturbations, among others. This may reflect in patients changing cluster assignments during the course of their disease, reflecting distinct symptomatic states over time. While the studies cited in this section do not particularly reflect that (with the exception of Geisler et al), there is a growing interest in analysing longitudinal symptom clusters, not only to dissect putative causes of disease but also with the aim of improving prognosis by building forecasting models [51].

The second set of articles (table 2) focused on a very different approach. While, as discussed before, symptom information can be extremely useful for subtyping patients due to the higher data availability and standardization of the recordings, the results do not necessarily correspond to biotypes and are not likely to be stable over time, reflecting more disease states rather than entities. When clustering biomarkers, however, the assumption is that the data capture the manifestation of the disorder at a lower level, yielding results that are potentially closer to uncovering pathological origins. To illustrate this claim with an example, there have been studies which show how different genetic alterations which produced different structural consequences, led to the same set of autistic-like behavioural traits in mice [52].

Among the methodologies overviewed in this section, structural MRI clustering and its projection into functional data deserve special mentioning. Several studies have shown that psychiatry-related disorders have in many cases structural implications. While diseases such as Autism or Schizophrenia are generally recognised as neurodevelopmental disorders with brain structure being affected, there are inconsistencies regarding the regional specificity of the neuroanatomical findings, which make apparent the importance of structural subtyping. Furthermore, the search for functional correlates of these subtype-specific functional alterations relies on the assumption that an altered structure may lead to an altered function. By combining the two data types, it is possible to test this hypothesis, retrieving multiple domains affected by the disorder that may be coupled with a non-trivial causal relationship.

The third and last set of articles (table 3) dealt with unsupervised learning on functional MRI itself, a technique with the potential of shedding light into mechanistic biotypes reflecting distinct pathological entities that overlap at higher levels. While, on the interpretation side, this category is not substantially different from the previous one, the distinction was made based on the usage that the researchers gave to the functional MRI data itself. Both task and resting-state approaches have been explored, although the vast majority of studies opted for the latter given its easier implementation and potentially broader conclusions and generalisability. The employed pipelines were extremely variable, using both functional and effective connectivity and a broad collection of clustering algorithms, the simple K-means and graph theory-based methods being the favourite tools.

Regarding testing of the presence of a cluster structure against a continuum in all three categories, several studies overlooked this issue completely, and most included K=1 as a possible cluster solution during model selection. Very few included dedicated statistical tests to evaluate the presence of clusters alone. Stability, however, was thoroughly tested in most of the articles, leave-one-out cross-validation and random network permutations (for graph-based systems) being the most employed methodologies. The sample sizes, however, remain mostly small and much more testing and replication is needed, especially concerning functional biotypes.

In this regard, as MRI data has an extremely high number of features, dimensionality reduction techniques are nearly mandatory when the sample sizes are small. While there are several manners of narrowing the feature space, special mention goes to the ways of integrating external domain knowledge into the functional clustering pipelines, either by directed dimensionality reduction (using CCA or its related variants[53], for example) or by the usage of meta-analytic masks for feature selection (such as the ones provided by NeuroSynth). The former can be handy to gather functional correlates of expressed symptoms but has the downside of shifting the attention from the pure mechanistic biotype elucidation, most likely yielding clusters based on functional manifestations of phenomena that can be observed at a higher level. In any case, proper statistical testing has to be carried out to avoid overfitting, to which several algorithms in this family are prone [54]. Meta-analyses, for their part, are a simple way of selecting features that have already been implicated in the disease at hand. While this can be useful to reduce the feature space and extract potentially more meaningful conclusions, the study may lose information about brain areas not yet reported to be relevant, for example by being implicated in only a subset of the possible biotypes.

To conclude, although functional MRI has a great potential in several areas of precision medicine such as diagnosis, prognosis and treatment selection, the results achieved so far are inconsistent in all cases, and much remains to be done in terms of gathering and centralising data, standardising pipelines and model validation, and method refinement. Furthermore, as most of the presented studies started from symptom-defined broad labels (such as Major Depression or ADHD, for example), much remains to be explored in the field of trans-diagnostic data-driven psychiatry [55]. The potential contribution of functional MRI to this field, that as mentioned in the introduction searches for a data-driven replacement for the current canon DSM labels, remains vastly uncharted.